\documentclass[pdflatex,sn-nature]{sn-jnl}

\usepackage{graphicx}%
\usepackage{multirow}%
\usepackage{amsmath,amssymb,amsfonts}%
\usepackage{amsthm}%
\usepackage{mathrsfs}%
\usepackage[title]{appendix}%
\usepackage{xcolor}%
\usepackage{textcomp}%
\usepackage{manyfoot}%
\usepackage{booktabs}%
\usepackage{algorithm}%
\usepackage{algorithmicx}%
\usepackage{algpseudocode}%
\usepackage{listings}%
\usepackage{verbatim} %
\usepackage{float} %
\usepackage{bm} 

\theoremstyle{thmstyleone}%

\theoremstyle{thmstyletwo}%

\theoremstyle{thmstylethree}%

\begin{document}
\newgeometry{left=28mm,right=28mm,top=28mm,bottom=28mm}

\title[Code-space recovery]{Code-space recovery for sample-based quantum diagonalization beyond native symmetry constraints}

\author[1]{\fnm{Byeongyong} \sur{Park}}\email{by7816@uos.ac.kr}

\author[1]{\fnm{Sanha} \sur{Kang}}\email{sanha9156@uos.ac.kr}

\author[1,2]{\fnm{Doyeol (David)} \sur{Ahn}}\email{dahn@uos.ac.kr}

\author*[3]{\fnm{Keunhong} \sur{Jeong}}\email{doas1mind@sogang.ac.kr}

\affil[1]{\orgdiv{Center for Quantum Information Processing}, \orgname{University of Seoul}, \orgaddress{\street{163 Seoulsiripdae-ro, Dongdaemun-gu}, \city{Seoul}, \postcode{02504}, \country{Republic of Korea}}}

\affil[2]{\orgname{Singularity Quantum Inc.}, \orgaddress{\street{9506 Villa Isle Drive}, \city{Villa Park, CA}, \postcode{92861}, \country{USA}}}

\affil[3]{\orgdiv{Department of Chemistry}, \orgname{Sogang University}, \orgaddress{\street{35 Baekbeom-ro, Mapo-gu}, \city{Seoul}, \postcode{04107}, \country{Republic of Korea}}}

\abstract{\unboldmath
Quantum algorithms that sample basis states and diagonalize a Hamiltonian in their span depend on repairing noisy measurement outcomes, using a constraint native to the target problem---typically particle-number symmetry. Many eigenvalue problems possess no such constraint.
Here we show that the constraint can instead be engineered into the sampling representation.
Using dual-rail encoding, \(\lvert0\rangle\to\lvert01\rangle\) and \(\lvert1\rangle\to\lvert10\rangle\), we implement sampling operations in encoded form, making code-space violations local recovery signals for self-consistent, reference-based repair. 
On transverse- and mixed-field Ising models of up to 36 spin sites, which lack a \(\mathrm{U}(1)\) symmetry usable for recovery, code-space recovery reached lower Ritz energies than diagonalization over the full observed support of matched unencoded samples while using \(38\text{--}84\%\) fewer basis states, despite a twofold qubit overhead. 
The margin was largest for the two-dimensional and 36-site systems and was already present in the first recovery iteration. 
Recoverable structure can be engineered rather than inherited.
}

\maketitle

\section{Introduction}\label{introduction}
Eigenvalue problems are fundamental computational tasks across science and engineering, arising in quantum many-body physics, electronic-structure theory, and data analysis~\cite{ev1}.
In many applications, the objective is to extract only selected spectral information, such as low-lying eigenvalues and their associated eigenspaces.
Nevertheless, reliably extracting these spectral features becomes increasingly demanding as the dimension of the underlying matrix grows, posing a major bottleneck in computational science~\cite{ev2}.

Quantum computers can manipulate states in Hilbert spaces whose dimension grows exponentially with qubit number, motivating the development of quantum algorithms to address large-scale eigenvalue problems~\cite{nc,qev1,qev2_ft1,qev3,qev4,qpe1,qpe2,vqe1,vqe2,vqe3,kqd}.
Quantum phase estimation is a canonical eigenvalue-estimation algorithm with rigorous guarantees on accuracy and success probability~\cite{nc,qpe1,qpe2}, but its circuit costs for practical problem sizes and target precisions generally place it beyond the reach of pre-fault-tolerant hardware~\cite{qev2_ft1,ft2,ft3}.
Variational quantum eigensolvers were introduced as hybrid alternatives~\cite{vqe1,vqe2,vqe3}, but their scalability can be limited by measurement costs, barren plateaus, and sensitivity to noise~\cite{bp1,bp2,bp3,mea_cost}.

Recently, several approaches have been proposed in which a quantum device serves as a sampler to identify relevant basis states whose span can provide a compact approximation to the target eigenspace~\cite{qsci,qsci_hci,sqd,skqd,sqdrift}.
A representative example is sample-based quantum diagonalization (SQD), originally developed for quantum chemistry~\cite{sqd}.
SQD samples candidate configurations on a quantum processor, stochastically repairs noisy samples using physical constraints such as particle-number symmetry, and classically diagonalizes the target Hamiltonian in the span of the recovered configurations.
This workflow has evolved into a broader quantum-centric framework through connections with Krylov subspace methods~\cite{skqd,sqdrift}, classical computational chemistry workflows~\cite{extsqd, afqmc, embedding, lsqd}, and machine-learning techniques~\cite{csqd, piegen}.
Applications to a molecule with half-M\"obius topology~\cite{science} and to protein--ligand complexes~\cite{protein_ligand} illustrate the expanding scope of SQD in realistic chemical and biomolecular modeling.

The SQD workflow is not intrinsically restricted to quantum chemistry.
In principle, the same strategy can be applied when the target eigenstates are sparse or compressible in a measurement basis accessible to the quantum sampler.
However, a key factor in the performance of existing SQD implementations is self-consistent recovery based on native constraints of the target problem.
At the sample level, this procedure identifies bitstrings that violate the relevant constraint and stochastically repairs them to satisfy it.
The importance of this step is evident in large-scale SQD demonstrations.
For example, in the \(\mathrm{[4Fe\!-\!4S]}\) experiment of Ref.~\cite{sqd}, the fraction of raw quantum samples lying in the correct particle-number sector was only on the order of \(10^{-5}\).
Thus, most raw samples could not be used directly as basis configurations in that sector, suggesting that recovery was critical to the performance of SQD.
This dependence on native constraints limits the practical applicability of the SQD workflow to more general eigenvalue problems that lack a native, measurement-visible constraint.

Here, we show that the sample-level constraint required for recovery need not be inherited from the target problem: it can be engineered into the representation in which the quantum sampler operates. 
We use the standard dual-rail encoding~\cite{dr1,dr2,dr3}, \(|0\rangle \mapsto |01\rangle\) and \(|1\rangle \mapsto |10\rangle\), not to protect logical states but to supply a recovery interface. 
The encoding assigns each logical qubit to a physical pair constrained to have Hamming weight one, so that 00 and 11 become directly identifiable code-space violations. 
Crucially, we implement logical sampling operations in encoded form, so that the encoded circuit reproduces the logical action within the code space throughout the sampling evolution.
Had the logical state been prepared first and the encoding applied only before measurement, error-corrupted outputs would themselves be mapped into the code space and pair validity would carry no information about accumulated errors. 
In noisy samples, code-space violations then identify where recovery is needed, while reference vectors initialized from the encoded measurement distribution and subsequently updated self-consistently guide the stochastic repair of invalid pairs. 
The repaired samples are decoded, and the resulting bitstrings construct the projected subspace for diagonalization. 
Because the constraint is carried by the encoding rather than by the target operator, the same interface applies wherever the sampling operations admit an encoded implementation.

Dual-rail encoding doubles the sampling register and increases circuit depth and gate count, so this strategy deliberately accepts greater exposure to hardware noise in exchange for a recovery signal. 
To test whether the exchange is favorable, we benchmarked the approach on one- and two-dimensional transverse-field Ising models (TFIMs) and one-dimensional mixed-field Ising models (MFIMs) with up to 36 spin sites. Because the transverse-field terms do not conserve Hamming weight in the computational basis, none of these Hamiltonians admits a \(\mathrm{U}(1)\) symmetry analogous to the particle-number symmetry used in conventional SQD recovery. 
Using matched logical sampling protocols, we compared the code-space recovery workflow with an unencoded baseline that diagonalized over the full observed support of the unencoded hardware samples---a baseline no subset of that support can improve upon. 
Across all benchmarks, code-space recovery yielded lower projected Ritz energies while using \(38\text{--}84\%\) fewer basis states, at a cost of roughly twofold increases in active-qubit and two-qubit-gate count. 
The margin was largest for the two-dimensional and 36-site systems, where the advantage was already present at the smallest subspace dimension tested.

Control experiments show that this performance does not follow from violation detection alone.
Repairing invalid pairs at random, without guidance from the encoded measurement distribution, failed to reproduce it in every tested setting, and often remained worse than the unencoded baseline itself. 
Under a common reference-guided expansion rule, moreover, expanding the basis from recovered encoded samples yielded lower Ritz energies than expanding from unencoded samples in all tested settings, indicating that the recovered logical support was the more effective starting point.

\section{Results} \label{results}
\subsection{Code-space recovery through encoded sampling} \label{results:encoded_sampling}

To estimate the lowest eigenpair of a general Hermitian target operator \(O\) using SQD, we employ the standard dual-rail encoding~\cite{dr1,dr2,dr3} as a recovery interface that provides an engineered sample-level constraint.
The corresponding isometry
\(V:\mathcal{H}_{\mathrm{tar}}=(\mathbb{C}^{2})^{\otimes n}\rightarrow
\mathcal{H}_{\mathrm{enc}}=(\mathbb{C}^{2})^{\otimes 2n}\)
is defined as
\begin{equation}
V =
\left(
|01\rangle\langle 0|
+
|10\rangle\langle 1|
\right)^{\otimes n}.
\label{eq:encoded_sqd_isometry}
\end{equation}
It satisfies \(V^{\dagger}V=I_{\mathcal{H}_{\mathrm{tar}}}\), and its image
\(\mathcal{C}=\operatorname{Im}(V)\subset\mathcal{H}_{\mathrm{enc}}\)
defines the code space.
An encoded bitstring
\(\widetilde{\mathbf{x}}=(\widetilde{x}_{1},\ldots,\widetilde{x}_{2n})\)
is code-valid if and only if
\begin{equation}
\widetilde{x}_{2i-1}+\widetilde{x}_{2i}=1,
\qquad
i=1,\ldots,n.
\label{eq:encoded_sqd_pair_constraint}
\end{equation}
Thus, \(01\) and \(10\) are valid pair outcomes, whereas \(00\) and \(11\) are invalid pair outcomes that provide sample-level recovery signals.

We replace each target-space operation \(A\) in the sampling circuits with an encoded operation \(A'\) satisfying
\begin{equation}
A'V = VA.
\label{eq:encoded_sqd_intertwining}
\end{equation}
For an encoded input \(V|\psi_0\rangle\), this relation ensures that, in the noiseless limit, the encoded circuit remains within \(\mathcal{C}\) and yields the same measurement distribution as the logical circuit after decoding.
If the logical state were prepared first and \(V\) applied only before measurement, error-corrupted logical outputs would be mapped into \(\mathcal{C}\), so pair validity would contain no information about errors accumulated during state preparation.
By contrast, using encoded operations satisfying Eq.~\eqref{eq:encoded_sqd_intertwining} throughout the sampling circuit allows errors that produce code-space leakage to be detected through invalid local pair outcomes.

\begin{figure}[!t]
\centering
\includegraphics[width=1.0\textwidth]{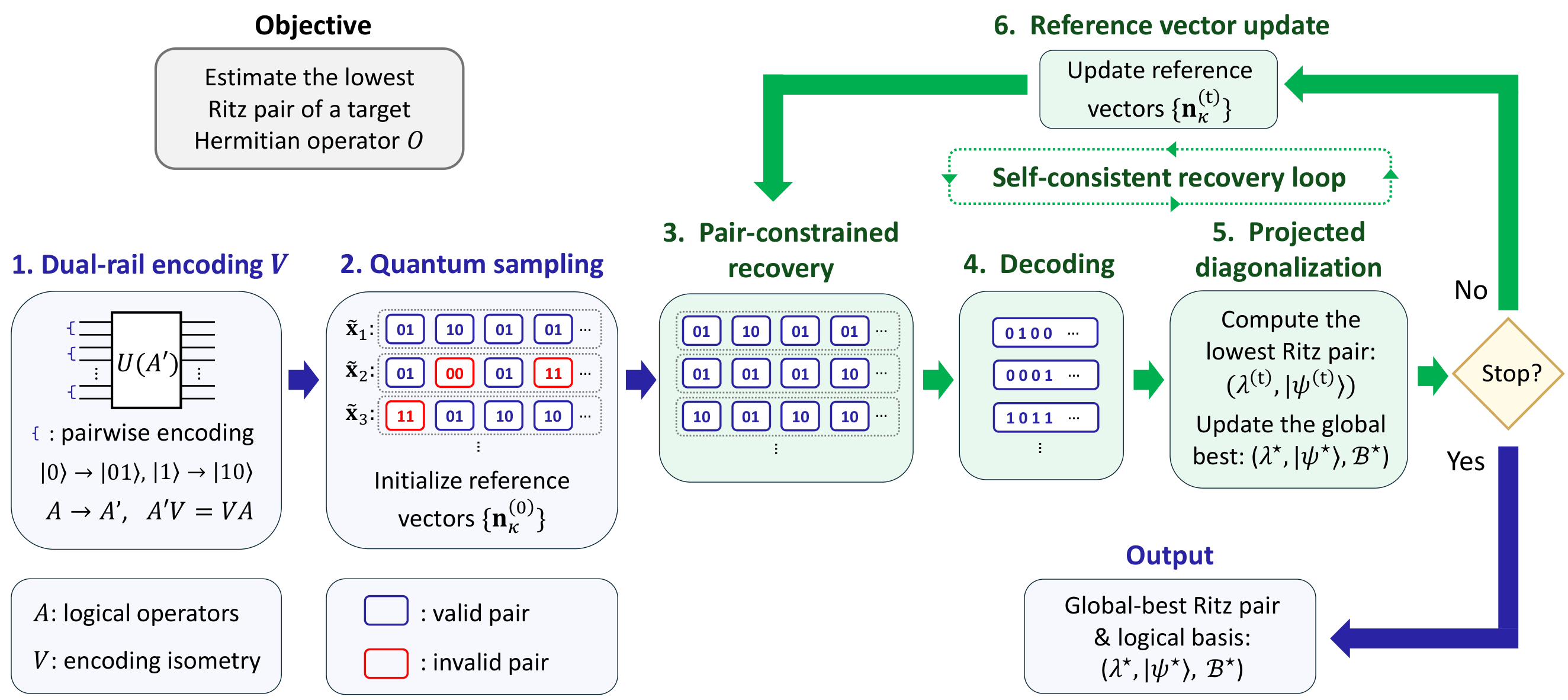}
\caption{\textbf{SQD workflow with code-space recovery.}
In sampling circuits, each logical operation \(A\) is replaced by an encoded operation \(A'\) satisfying \(A'V=VA\), where \(V\) is the dual-rail encoding isometry.
The encoded circuits produce \(2n\)-bit samples in which \(01\) and \(10\) are valid pair outcomes and \(00\) and \(11\) are detectable code-space violations.
The measured samples initialize reference vectors \(\{\mathbf{n}_{\kappa}^{(0)}\}\), which guide stochastic repair of invalid pairs.
Recovered strings are decoded and used to construct multiple candidate batches of logical basis states.
The target operator \(O\) is projected onto the span of each batch and diagonalized, and the batch yielding the lowest Ritz value defines the iteration-best result.
The global-best result is updated whenever the iteration-best Ritz value is lower than the current global-best value.
If the stopping condition is not met, the global-best Ritz state is used to update the reference vectors, which then guide recovery on the original measured encoded samples.
The workflow returns the global-best Ritz pair and logical basis \((\lambda^\star,|\psi^\star\rangle,\mathcal{B}^\star)\).}
\label{fig1}
\end{figure}

Fig.~\ref{fig1} summarizes the code-space recovery workflow.
The encoded circuits prepare \(V|\psi_0\rangle\), implement operations satisfying Eq.~\eqref{eq:encoded_sqd_intertwining}, and produce \(2n\)-bit hardware samples.
The self-consistent recovery procedure preserves valid pairs and stochastically repairs invalid pairs using reference vectors initialized from the measured encoded samples.
Each reference vector is a pair-normalized average rail-occupation profile.
To accommodate broad or multimodal support, we cluster the samples and maintain a separate reference vector for each cluster following the cluster-adaptive SQD strategy~\cite{csqd}.

The recovered code-valid bitstrings are decoded into logical bitstrings and used to construct multiple candidate basis batches with a prescribed target dimension \(D_{\mathrm{tar}}\).
The target operator \(O\) is projected and diagonalized in each batch, and the batch with the lowest Ritz value defines the iteration-best result.
This result replaces the global-best result when its Ritz value is lower.
The current global-best Ritz vector is then used to update the reference vectors, and logical basis states whose coefficient magnitudes meet the carry-over threshold are retained for the next iteration.
Details of the recovery algorithm and encoded operations are provided in Methods, with additional theoretical analysis in Supplementary Section~S1.

\subsection{Controlled encoded--unencoded benchmark design} \label{results:benchmark_setup}

We considered ground-state energy estimation for open-boundary Ising-type Hamiltonians on one-dimensional chains and two-dimensional square lattices of the form
\begin{equation}
H =
J\sum_{\langle i,j\rangle} Z_i Z_j
+
h_x\sum_i X_i
+
h_z\sum_i Z_i,
\label{eq:benchmark_ising_hamiltonian}
\end{equation}
where \(\langle i,j\rangle\) denotes nearest-neighbor pairs.
We set \(O=H\) and denote its Ritz values by \(E\), with the lowest providing a variational estimate of the ground-state energy.
We refer to models with \(h_z=0\) and \(h_z\neq 0\) as the TFIM and MFIM, respectively.
We set the ferromagnetic coupling to \(J=-1\) and report field strengths and energies in units of \(|J|\).
The one-dimensional models used \(h_x=-0.5\), with \(h_z=0\) for the TFIM and \(h_z=-0.5\) for the MFIM, whereas the two-dimensional TFIMs used \(h_x=-1\) and \(h_z=0\).
Our benchmark set consists of three \(n=25\) spin systems---the 1D TFIM, the 1D MFIM, and the \(5\times5\) 2D TFIM---and two \(n=36\) spin systems---the 1D MFIM and the \(6\times6\) 2D TFIM.

Because the transverse-field terms \(X_i\) flip individual spins, these Hamiltonians do not conserve total \(Z\)-magnetization, or equivalently Hamming weight in the computational \(Z\) basis.
Thus, unlike conventional electronic-structure SQD settings~\cite{sqd}, in which particle-number symmetry provides a native recovery constraint, these spin models lack an analogous \(\mathrm{U}(1)\) symmetry usable for sample recovery.

We generated quantum samples for basis construction with the SqDRIFT protocol~\cite{sqdrift}, which uses qDRIFT randomized compilation~\cite{qdrift} to approximate the time evolution defining the quantum Krylov states \(\lvert\psi_k\rangle=e^{-ikH\Delta t}\lvert\psi_0\rangle\).
We pooled hardware samples from \(k=1,2,3\).
The \(k=0\) Krylov state was the \(n\)-qubit Greenberger--Horne--Zeilinger (GHZ) state \(\lvert\psi_0\rangle\), whose computational-basis support consists of \(\lvert0^n\rangle\) and \(\lvert1^n\rangle\).
Rather than sampling this known support, we included these two logical bitstrings deterministically in every basis used for projected diagonalization.

The unencoded and encoded workflows used matched logical SqDRIFT circuits implemented directly on \(n\) qubits and through pair-code lifts on \(2n\) qubits, respectively.
This preserved the logical sampling protocol while retaining the physical overhead of encoding.
For each benchmark and workflow, we ran \(1000\) SqDRIFT circuits per Krylov index \(k\in\{1,2,3\}\) with \(1024\) shots per circuit on the IBM Heron r2 processor \texttt{ibm\_kingston}, yielding \(3.072\times10^6\) shots per pool.
The quantum runtime for each sample pool was approximately 15 min.
Both workflows used XpXm dynamical decoupling~\cite{dd} and matrix-free measurement mitigation (M3) with balanced calibration~\cite{m3}.
Across benchmarks, the encoded-to-unencoded ratios of median active-qubit count, transpiled depth, and two-qubit-gate count were \(1.97\text{--}2.18\), \(1.30\text{--}1.47\), and \(2.02\text{--}2.46\), respectively.
Details of the SqDRIFT sampling implementation are provided in Methods, with circuit-resource metrics reported in Supplementary Table~S2.

\subsection{Recovery-enhanced subspace quality} \label{results:benchmark_results}

To evaluate recovered-subspace quality, we compared code-space recovery with an unencoded baseline obtained by projecting and diagonalizing the logical Hamiltonian over the full observed support of the matched unencoded hardware sample pool.
By the variational principle, no subspace spanned by a subset of the observed unencoded support can yield a lower Ritz energy than this full-support baseline.
The unencoded full-support basis dimensions were approximately \((2.416\text{--}3.037)\times10^6\) across the five benchmarks.
For each benchmark, code-space recovery was evaluated at target basis dimensions \(D_{\rm tar}\in\{0.5,1.0,1.5,2.0,2.5\}\times10^6\) over five algorithmic seeds and 10 candidate batches per iteration using a fixed encoded hardware sample pool.
Here, \(D_{\rm tar}\) denotes the requested number of distinct logical basis states in each candidate batch, whereas \(D_{\rm proj}\) denotes the realized basis dimension used for projected diagonalization.
Across the five benchmarks, five target dimensions, and five algorithmic seeds, all 125 runs reached the requested target dimension, giving \(D_{\rm proj}=D_{\rm tar}\).

Final code-space recovery results were defined as the global-best Ritz energies obtained under the adaptive stopping rule described in Methods, subject to a 40-iteration cap.
Across the 125 runs, iteration counts ranged from 6 to 40 with a median of 13, and 124 runs met the stopping criterion while the remaining run reached the 40-iteration cap.

\begin{figure}[!t]
\centering
\includegraphics[width=1.0\textwidth]{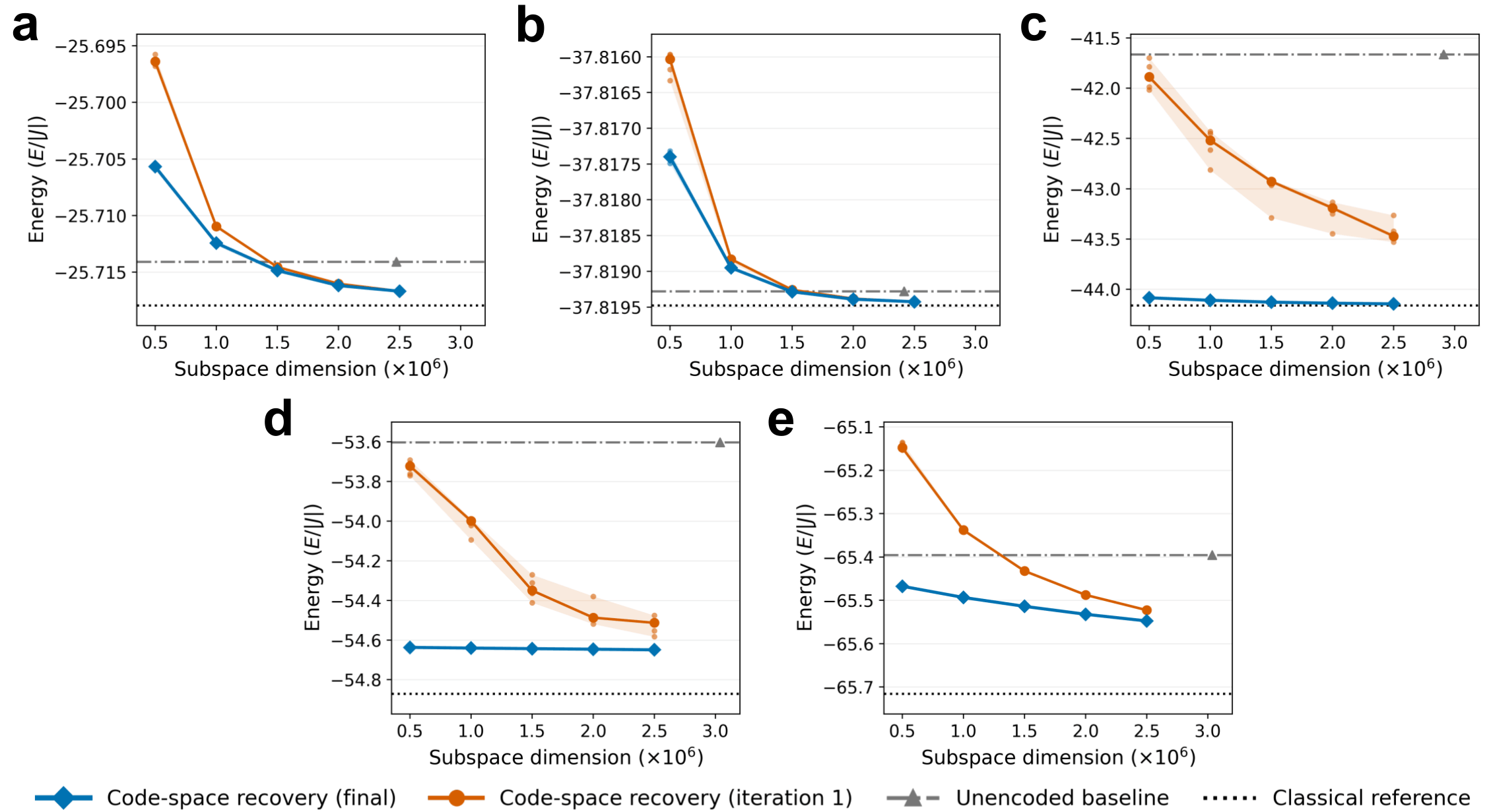}
\caption{\textbf{Ritz energies from code-space recovery and the unencoded baseline.}
Ritz energies obtained by applying code-space recovery to encoded hardware samples are plotted against the target subspace dimension and compared with the full-support baseline constructed from unencoded hardware samples and with classical reference energies.
Results are shown for \textbf{(a)} the 25-site 1D TFIM, \textbf{(b)} the 25-site 1D MFIM, \textbf{(c)} the \(5\times5\) 2D TFIM, \textbf{(d)} the 36-site 1D MFIM, and \textbf{(e)} the \(6\times6\) 2D TFIM.
Small points show individual seed results, while blue diamonds and orange circles connected by solid lines show final and first-iteration medians, respectively.
Shaded regions indicate min--max ranges.
Gray dash-dotted lines and triangles mark unencoded baseline energies and full-support basis sizes, respectively.
Black dotted lines indicate the classical reference energies, with the exact reference energy used for the 25-site 1D TFIM and density-matrix renormalization group (DMRG)~\cite{dmrg} reference energies used for the other four benchmarks.}
\label{fig2}
\end{figure}

The results are summarized in Fig.~\ref{fig2}.
For the 25-site 1D TFIM, the 25-site 1D MFIM, the \(5\times5\) 2D TFIM, the 36-site 1D MFIM, and the \(6\times6\) 2D TFIM, all five seeds yielded final Ritz energies below the corresponding unencoded full-support baselines for every tested \(D_{\rm tar}\) at or above the respective values of \((1.5,1.5,0.5,0.5,0.5)\times10^6\).
At the lowest such target dimensions, the recovered bases contained approximately \(39\%\), \(38\%\), \(83\%\), \(84\%\), and \(84\%\) fewer states than the corresponding unencoded full-support bases, respectively.

Notably, before any Ritz-state-based reference update or carry-over, all five seeds already yielded first-iteration Ritz energies below the corresponding unencoded baselines for every tested \(D_{\rm tar}\) at or above the respective values of \((1.5,2.0,0.5,0.5,1.5)\times10^6\) in the same benchmark order.
At the lowest such target dimensions, the recovered bases contained approximately \(39\%\), \(17\%\), \(83\%\), \(84\%\), and \(51\%\) fewer states than the corresponding unencoded full-support bases, respectively.

To probe robustness to hardware-sample variation, we partitioned each sample pool into two disjoint halves and compared code-space recovery with the full-support baselines constructed from the corresponding unencoded half-pools.
In both half-pools, code-space recovery could yield lower Ritz energies than the corresponding unencoded baselines while using smaller projected bases (Supplementary Section~S3.1).
A complementary no-clustering ablation using a single global reference showed that the two-cluster procedure used in the main experiments generally yielded lower Ritz energies.
Nevertheless, the single-reference control retained much of the advantage over the unencoded full-support baseline, indicating that clustering improved performance but was not essential (Supplementary Section~S3.2).

The lower-energy advantage of code-space recovery cannot be explained solely by the error-detection capability of dual-rail encoding.
Across the benchmarks, bitstrings in which all pairs were valid accounted for only \(0.0236\text{--}4.77\%\) of the probability mass in the M3-mitigated sample distributions.
Accordingly, strict postselection did not produce a competitive variational subspace (Supplementary Section~S2.2).
By contrast, the marginal probability that an individual pair was valid was \(74.4\%\text{--}89.4\%\), indicating that substantial local structure remained.

To test whether information in the encoded measurement distribution contributed to recovery performance beyond generic repair of invalid pairs, we used uniform random recovery as a control.
At each iteration, each invalid pair \(00\) or \(11\) in the encoded sample pool was mapped to \(01\) or \(10\) with equal probability \(1/2\), without reference-vector guidance.
The control used the same encoded hardware sample pool, target dimensions, five algorithmic seeds, 10 candidate batches per iteration, carry-over rule, and 40-iteration budget as the main code-space recovery experiments, but ran for all 40 iterations without early stopping.

\begin{figure}[!t]
\centering
\includegraphics[width=1.0\textwidth]{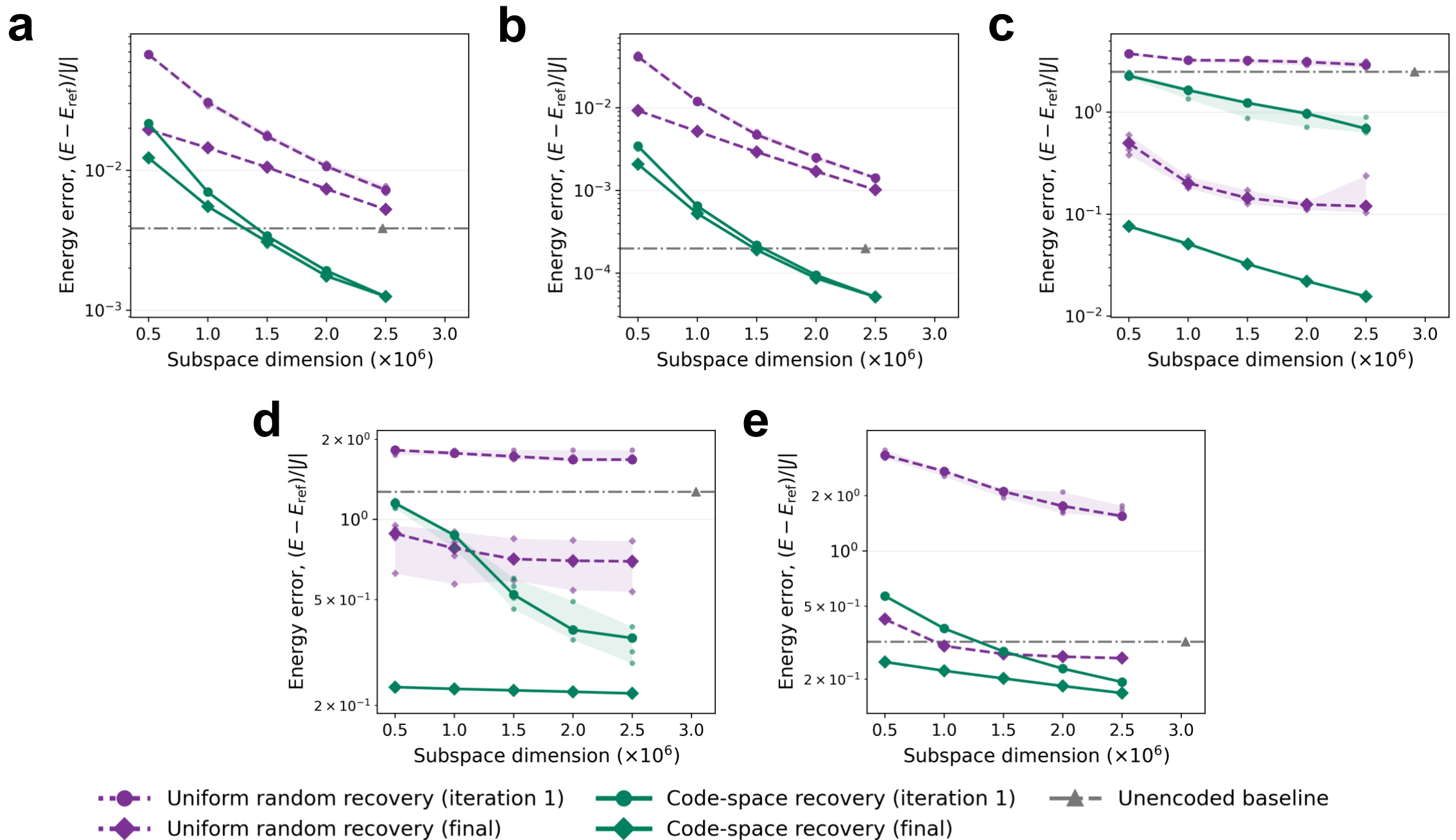}
\caption{\textbf{Ritz-energy errors for code-space recovery and uniform random recovery.}
Ritz-energy errors relative to the corresponding classical reference,
\((E-E_{\mathrm{ref}})/|J|\), are plotted on a logarithmic scale as a function of the target subspace dimension.
Here, \(E_{\mathrm{ref}}\) denotes the corresponding classical reference energy described in Fig.~\ref{fig2}.
Results are shown for \textbf{(a)} the 25-site 1D TFIM, \textbf{(b)} the 25-site 1D MFIM, \textbf{(c)} the \(5\times5\) 2D TFIM, \textbf{(d)} the 36-site 1D MFIM, and \textbf{(e)} the \(6\times6\) 2D TFIM.
Purple dashed and green solid lines show uniform random recovery and code-space recovery, respectively, while circles and diamonds mark first-iteration and final global-best results, respectively.
Final values were obtained under the stopping rule with a 40-iteration cap for code-space recovery and after exactly 40 iterations for uniform random recovery.
Small points, larger symbols connected by lines, and shaded regions show the five seed-level results, their medians, and min--max ranges, respectively.
Gray dash-dotted lines and triangles mark the errors and basis sizes of the unencoded full-support baselines, respectively.
}
\label{fig3}
\end{figure}

The results are summarized in Fig.~\ref{fig3}.
In the first iteration, the observed five-seed energy ranges of code-space recovery lay entirely below the corresponding uniform-random-recovery ranges in all 25 benchmark--dimension settings.
For every benchmark at \(D_{\rm tar}\geq2.0\times10^6\), the unencoded full-support baseline lay strictly between the two five-seed ranges, giving the Ritz-energy ordering uniform random recovery \(>\) unencoded full-support baseline \(>\) code-space recovery.

Repeated stochastic recovery and carry-over lowered the global-best energies of uniform random recovery.
However, in every benchmark--dimension setting, the lowest final energy obtained with uniform random recovery remained above the highest final energy obtained with code-space recovery.
Together, these results show that dual-rail violation detection, generic code-valid repair, iterative stochastic search, and coefficient-based carry-over alone did not reproduce the performance of code-space recovery.
The comparison further suggests that both the measurement-derived initialization of the reference vectors and their subsequent self-consistent updates contribute to the observed performance of code-space recovery.

\begin{figure}[!t]
\centering
\includegraphics[width=1.0\textwidth]{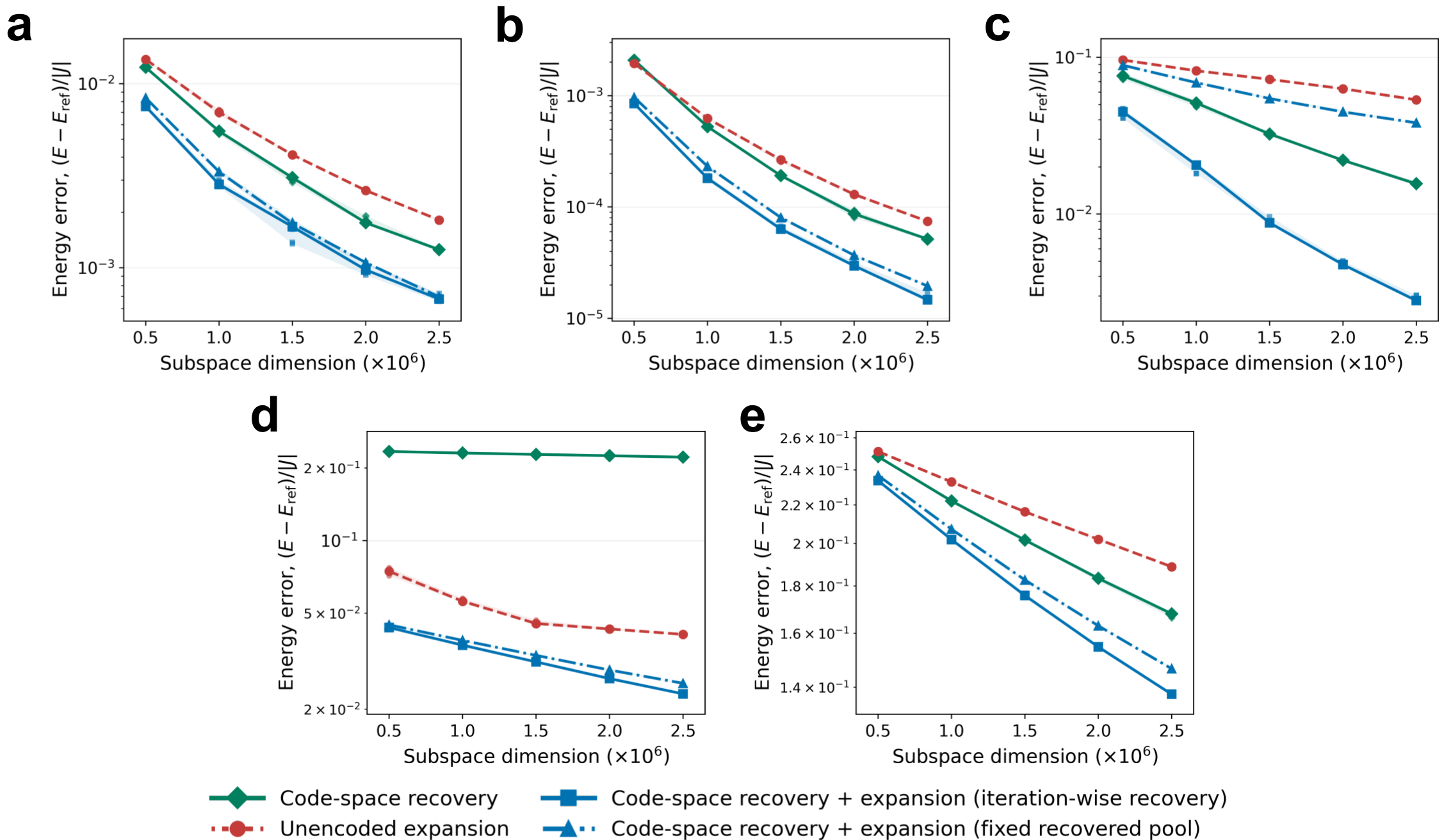}
\caption{\textbf{Ritz-energy errors for subspace expansion from encoded and unencoded samples.}
Ritz-energy errors relative to the corresponding classical reference,
\((E-E_{\mathrm{ref}})/|J|\), are plotted on a logarithmic scale versus the target subspace dimension.
Here, \(E_{\mathrm{ref}}\) denotes the corresponding classical reference energy described in Fig.~\ref{fig2}.
Results are shown for \textbf{(a)} the 25-site 1D TFIM, \textbf{(b)} the 25-site 1D MFIM, \textbf{(c)} the \(5\times5\) 2D TFIM, \textbf{(d)} the 36-site 1D MFIM, and \textbf{(e)} the \(6\times6\) 2D TFIM.
Small points, larger symbols connected by lines, and shaded regions show the five seed-level results, their medians, and min--max ranges, respectively.
Red dashed lines and circles show unencoded expansion using the observed unencoded sample pool as a fixed base pool.
Blue solid lines and squares show encoded expansion with iteration-wise recovery, which reapplied the code-space recovery rule to the original encoded samples at each iteration.
Blue dash-dotted lines and triangles show encoded expansion with a fixed recovered pool, which applied the rule once and reused the resulting logical pool.
Final values for the three expansion procedures are the global-best results obtained over exactly 40 iterations.
For comparison, green solid lines and diamonds show the final code-space recovery results without subspace expansion from Fig.~\ref{fig2}.}
\label{fig4}
\end{figure}

We next tested whether encoded samples provide a more effective source of basis states for subspace expansion by comparing three reference-guided procedures.
Unencoded expansion used the observed unencoded sample pool as a fixed base pool.
Encoded expansion with a fixed recovered pool applied the code-space recovery rule once at initialization to the original encoded samples and thereafter reused the resulting logical pool as a fixed base pool.
Encoded expansion with iteration-wise recovery reapplied the code-space recovery rule to the original encoded samples at each iteration using the current reference vectors, regenerating the base pool.
All three procedures generated reference-guided single- and double-bit-flip mutants from their respective base pools, updated the reference vectors self-consistently using the current global-best Ritz state, and applied coefficient-based carry-over.
They used the same five target dimensions and five algorithmic seeds, evaluated 10 candidate batches per iteration, and ran for exactly 40 iterations without early stopping.

Fig.~\ref{fig4} compares the three expansion procedures with the final results of code-space recovery without expansion shown in Fig.~\ref{fig2}.
In all 25 benchmark--dimension settings, the observed five-seed energy-error ranges of both encoded-expansion procedures lay entirely below the corresponding unencoded-expansion range.
In 19 of the 25 settings, the unencoded-expansion range lay entirely above even the range obtained by code-space recovery without expansion.
Encoded expansion with iteration-wise recovery yielded the lowest five-seed median energy error among all four methods in all 25 benchmark--dimension settings.
Moreover, its entire five-seed energy-error range lay below that of encoded expansion with a fixed recovered pool in 22 of the 25 settings.
Together, these results suggest that support recovered from encoded samples was more effective for subspace expansion than the observed unencoded support and that iteration-wise recovery yielded more favorable variational subspaces than reusing a fixed recovered pool.

As an auxiliary consistency check, we tracked the Ritz energy and full-\(H\) variance as the projected basis was enlarged while retaining all previously included states (Supplementary Section~S3.3).
For every benchmark--seed sequence, both quantities decreased monotonically with basis dimension, consistent with improved eigenstate character of the recovered Ritz states.
For the three 25-site benchmarks, linear energy-variance extrapolations~\cite{energy_variance} yielded zero-variance intercepts close to the corresponding classical reference energies, although this agreement does not strictly establish exact ground-state convergence.

Experimental and algorithmic details for the main experiments and controls are provided in Methods.
Detailed numerical results and classical wall-clock times for the main experiments are reported in Supplementary Section~S2.

\section{Discussion}\label{discussion}

In this work, we show that the recovery constraint underpinning sample-based diagonalization is a property of the sampling representation rather than of the target problem, and that it can therefore be engineered where it is absent. 
Dual-rail encoding supplies such a constraint by making code-space violations locally identifiable in noisy samples, at the cost of a doubled sampling register. 
Compared with full-support baselines constructed from matched unencoded hardware samples---baselines that no subset of the observed unencoded support can improve upon---code-space recovery yielded lower Ritz energies with smaller projected subspaces across all five benchmarks (Fig.~\ref{fig2}). 
This advantage was already evident in the first recovery iteration, before any Ritz-state-based reference update or carry-over, so it does not depend on extended iterative refinement; subsequent self-consistent refinement generally lowered the Ritz energies further.

The performance of code-space recovery cannot be explained solely by generic code-valid repair enabled by dual-rail violation detection, even when combined with iterative stochastic search and carry-over (Fig.~\ref{fig3}).
Before any Ritz-state-based reference update or carry-over, code-space recovery outperformed uniform random recovery in all tested settings.
Although iterative random recovery improved the control results, its final performance never matched that of code-space recovery.
Together, these results support an important role for recovery guidance provided by reference vectors initialized from the encoded measurement distribution and subsequently updated self-consistently.

The unencoded full-support baseline has a Ritz energy no higher than that obtained from any subset of the observed unencoded support.
Consequently, an unencoded workflow based on the corresponding sample pool can improve on this baseline only by expanding the basis beyond the observed support.
Using the same reference-guided expansion rule, procedures based on code-space recovery yielded lower Ritz energies than unencoded expansion in every tested setting (Fig.~\ref{fig4}).
These results suggest that the logical support obtained through code-space recovery was more effective for basis expansion than the observed unencoded support under the tested rule.
These comparisons do not establish that code-space recovery outperforms every unencoded basis-generation strategy.
Basis-expansion methods based on generative models~\cite{piegen} or Hamiltonian guidance~\cite{qsci_hci,extsqd} were not tested, and evaluating them in unencoded workflows and in combination with code-space recovery remains an important direction for future work.

In the code-space recovery procedure, dual-rail encoding is used not to implement quantum error correction but to engineer recoverable structure by making code-space violations identifiable in noisy samples.
A code-valid error that changes the logical value while remaining within the code space cannot be detected from the measurement outcome alone and is therefore not repaired by the recovery procedure.
Subspace-expansion methods may partially compensate for this limitation.
At the workflow level, dual-rail encoding increases quantum-resource requirements, while the iterative recovery procedure incurs classical computational overhead.
Accordingly, the present comparisons establish that code-space recovery can improve projected-subspace quality under the matched logical sampling protocols tested here, but not that it provides an end-to-end or resource-normalized computational advantage over unencoded workflows.

Although demonstrated here on Ising models, code-space recovery can in principle be extended to other Hermitian eigenvalue problems when the target low-lying eigenstate is sufficiently compressible in the chosen measurement basis and the sampling circuit assigns appreciable probability mass to the relevant support.
Such an extension requires an encoded implementation \(A'\) of each logical operation \(A\) satisfying \(A'V=VA\) for an encoding isometry \(V\), so that decoding reproduces the logical sampling distribution in the noiseless limit.
However, this relation constrains \(A'\) only within the code space.
The action of \(A'\) outside the code space and its circuit implementation are not uniquely determined and can affect circuit overhead, noise sensitivity, and code-space violation statistics.
Therefore, applying code-space recovery to other eigenvalue problems will require efficient design and optimization of these encoded implementations.

\section{Methods}\label{methods}

\subsection{Sampling protocol}

\paragraph{SqDRIFT protocol}
SqDRIFT~\cite{sqdrift} generates computational-basis samples from ensembles of randomized qDRIFT circuits~\cite{qdrift} that approximate the time evolutions defining quantum Krylov states~\cite{kqd, skqd}.
The target Hamiltonian is expressed as a linear combination of normalized Hermitian terms,
\begin{equation}
H
=
\sum_{\mu=1}^{M_h}
c_\mu h_\mu,
\qquad
c_\mu>0,
\qquad
\left\|h_\mu\right\|=1,
\label{eq:methods_sqdrift_hamiltonian_decomposition}
\end{equation}
where \(M_h\) is the number of terms and \(\|\cdot\|\) denotes the spectral norm.
Without loss of generality, coefficient signs and operator norms are absorbed into \(h_\mu\) and \(c_\mu\), respectively.

Let \(\lvert\psi_0\rangle\) denote the initial state.
For \(k=0,\ldots,K-1\), the propagation time is \(\tau_k=k\Delta t\), where \(\Delta t\) is the spacing between adjacent propagation times and \(K\) is the number of Krylov states.
The corresponding Krylov state is \(\lvert\psi_k\rangle=e^{-iH\tau_k}\lvert\psi_0\rangle\).
Under the ground-state sparsity and initial-overlap assumptions, the analysis in Ref.~\cite{skqd} establishes, for suitable \(K\) and \(\Delta t\), a lower bound on the probability of measuring each important ground-state bitstring in at least one Krylov state.

For each propagation time \(\tau_k\), SqDRIFT replaces the exact propagator \(e^{-iH\tau_k}\) with an ensemble of independently sampled qDRIFT realizations whose average channel approximates the corresponding exact unitary channel~\cite{qdrift}.
The coefficient \(\ell_1\)-norm and Hamiltonian-term sampling probabilities are defined as
\begin{equation}
\Lambda
=
\sum_{\mu=1}^{M_h}c_\mu,
\qquad
p_\mu
=
\frac{c_\mu}{\Lambda}.
\label{eq:methods_sqdrift_sampling_distribution}
\end{equation}
For each \(k=0,\ldots,K-1\), let \(r=0,\ldots,M_R-1\) index the qDRIFT realizations and \(m=1,\ldots,M_{\mathrm{seq}}\) index the sequence steps within each realization.
For every \((k,r,m)\), a Hamiltonian-term index \(\mu_{krm}\) is drawn independently with replacement according to \(\Pr(\mu_{krm}=\mu)=p_\mu\).

Given the sampled term sequence, the qDRIFT realization indexed by \(r\) for \(\tau_k\) and its constituent unitary factors are defined by
\begin{equation}
U_{kr}
=
W_{krM_{\mathrm{seq}}}
\cdots
W_{kr2}
W_{kr1},
\qquad
W_{krm}
=
\exp\left[
-i h_{\mu_{krm}}
\frac{\Lambda\tau_k}{M_{\mathrm{seq}}}
\right].
\label{eq:methods_sqdrift_randomized_unitary}
\end{equation}
The corresponding sampling circuit prepares \(\lvert\psi_0\rangle\) and applies \(U_{kr}\), producing the randomized sampling state \(\lvert\psi_{kr}^{\mathrm{qDRIFT}}\rangle=U_{kr}\lvert\psi_0\rangle\).

At each \(k\), the \(M_R\) realization circuits are measured \(M_S\) times in the computational basis, and the resulting \(M_RM_S\) bitstrings are pooled.
The pooled samples from the selected Krylov indices are then combined for basis construction and projected diagonalization.

\paragraph{Sampling implementation}
The benchmark Hamiltonians in Eq.~\eqref{eq:benchmark_ising_hamiltonian} are linear combinations of Pauli strings.
We therefore used each Pauli string as one normalized Hermitian term \(h_\mu\) in Eq.~\eqref{eq:methods_sqdrift_hamiltonian_decomposition}, absorbing the magnitude of its Hamiltonian coefficient into \(c_\mu\) and its sign into \(h_\mu\).

For all benchmarks, the logical initial state was chosen as the GHZ state
\begin{equation}
\lvert\psi_0\rangle
=
\frac{
\lvert 0^n\rangle
+
\lvert 1^n\rangle
}{
\sqrt{2}
},
\label{eq:methods_ghz_initial_state}
\end{equation}
since its two computational-basis components span the ground-state eigenspace of the benchmark Hamiltonians in the zero-field limit.
We used \(K=4\) Krylov states with a time spacing \(\Delta t=1.0\) and sampled only the nonzero indices \(k=1,2,3\) on quantum hardware.
Because the \(k=0\) Krylov state is the initial state itself, its logical computational-basis support consists exactly of \(\lvert0^n\rangle\) and \(\lvert1^n\rangle\).
We therefore did not execute a separate \(k=0\) circuit and included these states deterministically in every candidate basis used for projected diagonalization.

The encoded and unencoded workflows used matched logical SqDRIFT realizations.
At each sampled index \(k=1,2,3\), we generated \(M_R=1000\) independent logical qDRIFT realizations.
We set the qDRIFT sequence length to \(M_{\mathrm{seq}}=75\) for the 25-site 1D TFIM benchmark and \(M_{\mathrm{seq}}=100\) for all other benchmarks.
For each benchmark and pair \((k,r)\), the two workflows used the same ordered sequence of sampled logical Hamiltonian terms and the same evolution angles, differing only in the physical implementation of each logical realization.
In each workflow, every realization circuit was executed for \(M_S=1024\) shots.
Thus, each benchmark in each workflow used \(3M_R=3000\) circuits and produced \(3M_RM_S=3.072\times10^6\) measurement outcomes.

In the unencoded workflow, the state in Eq.~\eqref{eq:methods_ghz_initial_state} was prepared directly on an \(n\)-qubit register, and the randomized unitary \(U_{kr}\) in Eq.~\eqref{eq:methods_sqdrift_randomized_unitary} was implemented using the sampled Pauli-string evolutions.
Measurement of all \(n\) qubits in the computational basis yielded logical bitstrings \(\mathbf{x}=(x_1,\ldots,x_n)\in\{0,1\}^n\).

In the encoded workflow, the dual-rail isometry \(V\) defined in Eq.~\eqref{eq:encoded_sqd_isometry} mapped the initial state to
\begin{equation}
V\lvert\psi_0\rangle
=
\frac{
\lvert01\rangle^{\otimes n}
+
\lvert10\rangle^{\otimes n}
}{
\sqrt{2}
}.
\label{eq:methods_encoded_ghz_initial_state}
\end{equation}
At the circuit level, we initialized each second rail in \(\lvert1\rangle\), prepared the \(n\)-qubit GHZ state on the first rails, and applied a CNOT from each first rail to its paired second rail, implementing
\(\lvert x_i\rangle\lvert1\rangle\mapsto\lvert x_i,1-x_i\rangle\).

Because each normalized Hamiltonian term \(h_\mu\) is a signed Pauli string, we encoded its Pauli factors sitewise using
\begin{equation}
\begin{aligned}
I_i &\mapsto I_{2i-1}I_{2i}, \\
X_i &\mapsto X_{2i-1}X_{2i}, \\
Y_i &\mapsto Y_{2i-1}X_{2i}, \\
Z_i &\mapsto Z_{2i-1}I_{2i}.
\end{aligned}
\label{eq:methods_pauli_lift}
\end{equation}
We denote the resulting encoded Hamiltonian term by \(h'_\mu\).
This construction satisfies the intertwining condition in Eq.~\eqref{eq:encoded_sqd_intertwining}, with \(A=h_\mu\) and \(A'=h'_\mu\).

By the power-series definition of the matrix exponential, Eq.~\eqref{eq:encoded_sqd_intertwining} also implies \(e^{-i\theta h'_\mu}V=Ve^{-i\theta h_\mu}\) for any \(\theta\in\mathbb{R}\).
The encoded qDRIFT steps and realizations were defined by
\begin{equation}
W'_{krm}
=
\exp\left[
-i h'_{\mu_{krm}}
\frac{\Lambda\tau_k}{M_{\mathrm{seq}}}
\right],
\qquad
U'_{kr}
=
W'_{krM_{\mathrm{seq}}}
\cdots
W'_{kr2}
W'_{kr1}.
\label{eq:methods_encoded_sqdrift_realization}
\end{equation}
The resulting operators satisfy
\(W'_{krm}V=VW_{krm}\) and \(U'_{kr}V=VU_{kr}\), as required by the intertwining condition in Eq.~\eqref{eq:encoded_sqd_intertwining}.
Thus, in the absence of noise, the encoded and unencoded circuits implement the same logical SqDRIFT realization.
After the encoded evolution, measurement of all \(2n\) physical qubits in the computational basis yielded encoded bitstrings \(\widetilde{\mathbf{x}}=(\widetilde{x}_1,\ldots,\widetilde{x}_{2n})\in\{0,1\}^{2n}\).

All circuits were transpiled with Qiskit~\cite{qiskit} at optimization level \(3\) for the IBM Heron r2 processor \texttt{ibm\_kingston}.
Both workflows used XpXm dynamical decoupling~\cite{dd} and M3 measurement mitigation~\cite{m3} with balanced calibration.
Each M3 quasi-probability distribution was projected onto the probability simplex, and the resulting normalized nonnegative probabilities were used as sample weights.

\subsection{Code-space recovery}

In this section, we present a self-consistent recovery algorithm that takes a weighted set of noisy encoded bitstrings as input and returns the lowest Ritz pair found for a target Hermitian operator \(O\) acting on the logical space.
We assume that, in the noiseless limit, the sampling circuits assign sufficient probability mass to the encoded images of the logical computational-basis states that constitute the dominant support of a lowest-eigenvalue eigenstate of \(O\).
Apart from this sampling requirement, the recovery procedure is independent of the method used to generate the encoded bitstrings.

\paragraph{Input samples and encoding maps}
The input sample set is written as
\begin{equation}
\widetilde{\mathcal X}^{(0)}
=
\left\{
\left(
\widetilde{\mathbf{x}}^{(0)}_s,
w^{(0)}_s
\right)
\right\}_{s=1}^{N_{\rm uniq}^{(0)}},
\qquad
w_s^{(0)}>0,
\qquad
\sum_{s=1}^{N_{\rm uniq}^{(0)}}w_s^{(0)}=1,
\label{eq:methods_recovery_initial_weighted_samples}
\end{equation}
where \(N_{\rm uniq}^{(0)}\) denotes the number of unique encoded bitstrings and \(w_s^{(0)}\) denotes the weight assigned to bitstring \(\widetilde{\mathbf{x}}_s^{(0)}\).

For a logical bitstring
\(
\mathbf{x}=(x_1,\ldots,x_n)\in\{0,1\}^n
\),
the classical encoding and decoding maps of the pair code are defined by
\begin{equation}
\operatorname{Enc}(\mathbf{x})
=
\left(
x_1,1-x_1,\ldots,x_n,1-x_n
\right),
\qquad
\operatorname{Dec}(\widetilde{\mathbf{x}})
=
\left(
\widetilde{x}_1,
\widetilde{x}_3,
\ldots,
\widetilde{x}_{2n-1}
\right).
\label{eq:methods_recovery_encoding_maps}
\end{equation}
The decoding map is applied only to encoded bitstrings that satisfy the pair constraint in Eq.~\eqref{eq:encoded_sqd_pair_constraint}.

\paragraph{Initial clustering and reference vectors}

The initial weighted samples are partitioned into \(N_C\) clusters in the physical \(2n\)-bit encoded space using a weighted clustering method suitable for binary data, such as a Bernoulli mixture model (BMM)~\cite{bmm}.
Clustering uses the sample weights \(w_s^{(0)}\) without imposing the pair constraint.
Let \(g_s^{(0)}\in\{1,\ldots,N_C\}\) denote the initial cluster label of sample \(s\), and let \(\kappa\) index the clusters.

Each cluster is represented by a pair-normalized reference vector
\begin{equation}
\mathbf n_{\kappa}^{(t)}
=
\left(
n_{\kappa,1}^{(t)},
\ldots,
n_{\kappa,2n}^{(t)}
\right)
\in
[0,1]^{2n},
\label{eq:methods_recovery_reference_vector}
\end{equation}
where \(t=0\) denotes initialization and \(t\geq1\) denotes a recovery iteration.
For each encoded pair \(i=1,\ldots,n\), the initial reference vector is obtained from the weighted rail occupancies within cluster \(\kappa\) as
\begin{equation}
\begin{aligned}
n_{\kappa,2i-1}^{(0)}
&=
\frac{
\displaystyle
\sum_{s:g_s^{(0)}=\kappa}
w_s^{(0)}
\widetilde{x}_{s,2i-1}^{(0)}
}{
\displaystyle
\sum_{s:g_s^{(0)}=\kappa}
w_s^{(0)}
\left(
\widetilde{x}_{s,2i-1}^{(0)}
+
\widetilde{x}_{s,2i}^{(0)}
\right)
},
\\
n_{\kappa,2i}^{(0)}
&=
\frac{
\displaystyle
\sum_{s:g_s^{(0)}=\kappa}
w_s^{(0)}
\widetilde{x}_{s,2i}^{(0)}
}{
\displaystyle
\sum_{s:g_s^{(0)}=\kappa}
w_s^{(0)}
\left(
\widetilde{x}_{s,2i-1}^{(0)}
+
\widetilde{x}_{s,2i}^{(0)}
\right)
}.
\end{aligned}
\label{eq:methods_recovery_initial_reference}
\end{equation}
If the common denominator is zero, we set
\(
(n_{\kappa,2i-1}^{(0)},n_{\kappa,2i}^{(0)})=(1/2,1/2)
\).
Consequently, every reference pair satisfies \(n_{\kappa,2i-1}^{(0)}+n_{\kappa,2i}^{(0)}=1\).

\paragraph{Stochastic pairwise recovery}

Given an encoded bitstring
\(
\widetilde{\mathbf{x}}\in\{0,1\}^{2n}
\)
and a pair-normalized reference vector \(\mathbf n\), the stochastic recovery map
\(
R(\widetilde{\mathbf{x}};\mathbf n)
\)
leaves every valid pair unchanged and repairs each invalid pair by flipping the bit on one of its two rails.

We adopt the modified-ReLU rail-flip scoring rule and parameter prescription of SQD~\cite{sqd}.
The modified-ReLU function is defined as
\begin{equation}
\phi_{\rho,\delta}(u)
=
\begin{cases}
\displaystyle
\delta\frac{u}{\rho},
&
u\le\rho,
\\[8pt]
\displaystyle
\delta
+
(1-\delta)
\frac{u-\rho}{1-\rho},
&
u>\rho.
\end{cases}
\label{eq:methods_recovery_modified_relu}
\end{equation}
For the dual-rail pair code, we fixed \((\rho,\delta)=(1/2,0.01)\) throughout all benchmark experiments.

Consider the \(i\)th encoded pair
\(
(\widetilde{x}_{2i-1},\widetilde{x}_{2i})
\).
An invalid pair is either \(00\) or \(11\), and either rail can be flipped to produce a valid pair.
For each candidate rail
\(
j\in\{2i-1,2i\}
\),
we define the score
\begin{equation}
\Gamma_j
=
\phi_{\rho,\delta}
\left(
\left|
\widetilde{x}_j-n_j
\right|
\right).
\label{eq:methods_recovery_flip_score}
\end{equation}
The selected rail \(j\) is sampled with probabilities
\begin{equation}
\pi_{2i-1}
=
\frac{
\Gamma_{2i-1}
}{
\Gamma_{2i-1}+\Gamma_{2i}
},
\qquad
\pi_{2i}
=
\frac{
\Gamma_{2i}
}{
\Gamma_{2i-1}+\Gamma_{2i}
}.
\label{eq:methods_recovery_flip_probabilities}
\end{equation}
The bit on the selected rail is flipped.
The rail choices are sampled independently across invalid pairs.

\paragraph{Iterative recovery}

Recovery iterations are indexed by \(t=1,\ldots,N_{I,\max}\).
At every iteration, recovery is reapplied to the initial weighted sample set
\(
\widetilde{\mathcal X}^{(0)}
\).
The first iteration retains the initial cluster labels, so
\(
g_s^{(1)}=g_s^{(0)}
\).
For \(t\ge2\), each initial encoded bitstring is reassigned to the nearest reference vector from the preceding iteration in Manhattan distance according to
\begin{equation}
g_s^{(t)}
=
\operatorname*{arg\,min}_{\kappa=1,\ldots,N_C}
\left\|
\widetilde{\mathbf{x}}_s^{(0)}
-
\mathbf n_{\kappa}^{(t-1)}
\right\|_1.
\label{eq:methods_recovery_sample_reassignment}
\end{equation}

At iteration \(t\), every unique original bitstring \(\widetilde{\mathbf{x}}_s^{(0)}\) is processed once in an initial recovery draw.
Pair-valid bitstrings are retained, whereas invalid ones are mapped to
\(R\!\left(\widetilde{\mathbf{x}}_s^{(0)};\mathbf n_{g_s^{(t)}}^{(t-1)}\right)\), with each output inheriting the corresponding input weight.
If fewer than \(D_{\rm tar}\) distinct logical bitstrings are available under the batch-construction rules described below, the recovery map is independently reapplied to \(\widetilde{\mathcal X}^{(0)}\) with the cluster assignments and reference vectors held fixed.
Drawing continues until at least \(D_{\rm tar}\) eligible unique logical bitstrings are available or \(N_{\rm draw}^{\max}\) draws have been completed.
Outputs from the \(N_{\rm draw}^{(t)}\) completed draws are pooled, duplicates are merged, and their accumulated weights are divided by \(N_{\rm draw}^{(t)}\).

The resulting recovered sample set \(\widetilde{\mathcal X}^{(t)}\) is written as
\begin{equation}
\widetilde{\mathcal X}^{(t)}
=
\left\{
\left(
\widetilde{\mathbf{x}}^{(t)}_s,
w^{(t)}_s
\right)
\right\}_{s=1}^{N_{\rm uniq}^{(t)}},
\qquad
w_s^{(t)}>0,
\qquad
\sum_{s=1}^{N_{\rm uniq}^{(t)}}w_s^{(t)}=1,
\label{eq:methods_recovery_weighted_recovered_samples}
\end{equation}
where \(N_{\rm uniq}^{(t)}\) and \(w_s^{(t)}\) denote the number of unique recovered encoded bitstrings and the weight of \(\widetilde{\mathbf{x}}_s^{(t)}\), respectively.
By construction, every bitstring in \(\widetilde{\mathcal X}^{(t)}\) satisfies the pair constraint.

\paragraph{Decoding, batch construction, and projected diagonalization}

After recovery, valid encoded bitstrings are decoded into logical bitstrings, and all subsequent basis construction and projected diagonalization are performed in the logical space.
At every iteration, the algorithm constructs \(N_B\) logical basis batches.
Each batch is filled until it reaches the target size \(D_{\rm tar}\) or no additional eligible logical bitstrings remain.
Thus, \(D_{\rm tar}\) is an upper bound on each realized batch size and hence on the final projected dimension \(D_{\rm proj}\).
Let
\(
\mathcal B_{\rm fix}\subseteq\{0,1\}^n
\)
denote an optional fixed set included in every batch and
\(
\mathcal B_{\rm carry}^{(t)}
\)
the carry-over set produced after iteration \(t\), with
\(
\mathcal B_{\rm carry}^{(0)}=\emptyset
\).
At iteration \(t\), each batch is initialized as
\(
\mathcal B_b^{(t)}
\leftarrow
\mathcal B_{\rm fix}
\cup
\mathcal B_{\rm carry}^{(t-1)}
\)
for \(b=1,\ldots,N_B\).

After initialization, each batch is filled from an iteration-dependent candidate pool.
At \(t=1\), logical bitstrings decoded from the pair-valid members of \(\widetilde{\mathcal X}^{(0)}\) are prioritized.
If the eligible members of this set suffice to fill the remaining capacity, candidates are sampled only from this set.
Otherwise, all eligible members are included and the remaining slots are filled from logical bitstrings decoded from \(\widetilde{\mathcal X}^{(1)}\).
For \(t\geq2\), no separate priority is given to the pair-valid members of \(\widetilde{\mathcal X}^{(0)}\). After \(\mathcal B_{\rm fix}\) and \(\mathcal B_{\rm carry}^{(t-1)}\) are included, all remaining slots are filled from logical bitstrings decoded from \(\widetilde{\mathcal X}^{(t)}\).
Sampling is performed without replacement with probabilities proportional to the weights of the remaining candidates, independently for each batch, until the batch reaches \(D_{\rm tar}\) or no eligible unique candidates remain.

For each batch, the target operator is projected onto the corresponding logical subspace according to
\begin{equation}
P_{\mathcal B_b^{(t)}}
=
\sum_{\mathbf{x}\in\mathcal B_b^{(t)}}
\lvert\mathbf{x}\rangle
\langle\mathbf{x}\rvert,
\qquad
O_b^{(t)}
=
P_{\mathcal B_b^{(t)}}
O
P_{\mathcal B_b^{(t)}}.
\label{eq:methods_recovery_projected_operator}
\end{equation}
The normalized lowest Ritz vector within this subspace satisfies
\begin{equation}
O_b^{(t)}
\lvert\psi_b^{(t)}\rangle
=
\lambda_b^{(t)}
\lvert\psi_b^{(t)}\rangle,
\qquad
\lvert\psi_b^{(t)}\rangle
\in
\operatorname{ran}\!\left(P_{\mathcal B_b^{(t)}}\right).
\label{eq:methods_recovery_projected_eigenproblem}
\end{equation}
Among the \(N_B\) batches, the batch with the smallest \(\lambda_b^{(t)}\) defines the iteration-best result
\((\lambda^{(t)},\lvert\psi^{(t)}\rangle,\mathcal B^{(t)})\).

\paragraph{Carry-over and reference updates}

After the first iteration, we set
\(
\lambda^\star=\lambda^{(1)}
\),
\(
\lvert\psi^\star\rangle=\lvert\psi^{(1)}\rangle
\),
and
\(
\mathcal B^\star=\mathcal B^{(1)}
\).
For \(t\ge2\), these quantities are replaced by the iteration-best result whenever
\(
\lambda^{(t)}<\lambda^\star
\).
The global-best state is expanded as \(\lvert\psi^\star\rangle=\sum_{\mathbf{x}\in\mathcal B^\star}\alpha_{\mathbf{x}}^\star\lvert\mathbf{x}\rangle\).
Given a carry-over threshold \(\epsilon_{\rm carry}\), the carry-over set after iteration \(t\) is defined by
\begin{equation}
\mathcal B_{\rm carry}^{(t)}
=
\left\{
\mathbf{x}\in\mathcal B^\star
\;\middle|\;
\left|
\alpha_{\mathbf{x}}^\star
\right|
\ge
\epsilon_{\rm carry}
\right\}.
\label{eq:methods_recovery_carry_set}
\end{equation}
This set is included deterministically in every batch at the next iteration.

The cluster-specific reference vectors are updated using the current global-best state.
Let \(g_{\mathbf{x}}^{(t)}\) denote the cluster assigned to
\(\operatorname{Enc}(\mathbf{x})\) by its Manhattan distance to the previous reference vectors, as in Eq.~\eqref{eq:methods_recovery_sample_reassignment}.
The reference vector for cluster \(\kappa\) is updated as
\begin{equation}
\mathbf n_{\kappa}^{(t)}
=
\frac{
\displaystyle
\sum_{\substack{
\mathbf{x}\in\mathcal B^\star\\
g_{\mathbf{x}}^{(t)}=\kappa
}}
\left|
\alpha_{\mathbf{x}}^\star
\right|^2
\operatorname{Enc}(\mathbf{x})
}{
\displaystyle
\sum_{\substack{
\mathbf{x}\in\mathcal B^\star\\
g_{\mathbf{x}}^{(t)}=\kappa
}}
\left|
\alpha_{\mathbf{x}}^\star
\right|^2
}.
\label{eq:methods_recovery_reference_update}
\end{equation}
A cluster with no assigned weight retains its previous reference vector.

\paragraph{Stopping criterion}

After at least \(N_{I,\min}\) iterations, the loop terminates if the global-best Ritz value has not strictly improved for \(N_{\rm conv}\) consecutive iterations.
If this condition is not met, the loop terminates after \(N_{I,\max}\) iterations.
The final output is
\begin{equation}
\left(
\lambda^\star,
\lvert\psi^\star\rangle,
\mathcal B^\star
\right),
\qquad
D_{\rm proj}
=
\left|
\mathcal B^\star
\right|
\le
D_{\rm tar}.
\label{eq:methods_recovery_final_output}
\end{equation}

\subsection{Experimental settings}

\paragraph{Benchmark settings}
In the main code-space recovery sweep, one encoded hardware sample pool was used per benchmark and reused across the five target dimensions and five algorithmic seeds.
The seeds controlled clustering, stochastic recovery, and batch construction and did not represent independent hardware sampling.
The encoded samples were clustered using a weighted BMM~\cite{bmm} with \(N_C=2\).
Among 50 independent initializations, we retained the fit with the highest final likelihood lower bound and assigned each bitstring to one cluster.
We used
\(D_{\rm tar}\in\{0.5,1.0,1.5,2.0,2.5\}\times10^6\),
\(N_B=10\),
\(\epsilon_{\rm carry}=10^{-3}\),
\(N_{I,\min}=3\),
\(N_{I,\max}=40\),
\(N_{\rm conv}=5\), and
\(N_{\rm draw}^{\max}=10\).
For both code-space recovery and the unencoded baseline, every projected basis included
\(\mathcal B_{\rm fix}=\{0^n,1^n\}\).
The unencoded baseline was obtained from a single diagonalization of the original logical Hamiltonian in the augmented full-support basis containing \(\mathcal B_{\rm fix}\) and all unique logical bitstrings from the matched unencoded sample pool.

For each selected basis, we computed the lowest Ritz pair of the projected logical Hamiltonian using the Hermitian \texttt{eigsh} interface of PRIMME~\cite{primme}, with a tolerance of \(10^{-10}\) and the default method selection.

\paragraph{Classical reference energies}

For the 25-site 1D TFIM, the classical reference energy \(E_{\rm ref}\) was computed exactly using an open-boundary free-fermion calculation~\cite{1dtfim}.
For the other four benchmarks, \(E_{\rm ref}\) was obtained using the two-site DMRG implementation in quimb~\cite{dmrg,quimb}, with bond dimensions up to \(512\), a truncation cutoff and energy-convergence tolerance of \(10^{-10}\), and at most \(24\) sweeps.
The resulting values are reported in Supplementary Table~S4.

\paragraph{Control experiments}

Uniform random recovery and the two encoded expansion controls used the same encoded hardware sample pools as in the main code-space recovery experiments, whereas unencoded expansion used the matched unencoded pools.
Uniform random recovery treated each encoded pool as a single weighted pool and, at every iteration, reapplied uniform recovery to the original samples.
Valid pairs were retained, whereas each invalid pair was independently mapped to \(01\) or \(10\) with equal probability.
Additional independent recovery draws were pooled when needed to fill \(D_{\rm tar}\), up to \(N_{\rm draw}^{\max}=10\) per iteration.

The three expansion controls generated reference-guided single- and double-bit-flip mutants from their respective base pools to expand the candidate support.
Unencoded expansion used the observed unencoded support as a fixed base pool.
Encoded expansion with iteration-wise recovery regenerated its logical base pool at every iteration by applying one reference-guided pair-recovery pass to the original encoded samples.
Encoded expansion with a fixed recovered pool applied the same recovery once in the first iteration and reused the resulting logical pool thereafter.

The three expansion controls applied the weighted BMM procedure to the corresponding encoded or unencoded sample pools to obtain \(N_C=2\) initial clusters.
The two encoded-expansion controls used the pair-normalized references \(\mathbf n_\kappa^{(t)}\) defined above.
For unencoded expansion, each cluster was represented by a logical reference vector \(\mathbf r_\kappa^{(t)}\in[0,1]^n\).
Initially, \(\mathbf r_\kappa^{(0)}\) was the sample-weighted mean of the logical bitstrings in cluster \(\kappa\).
For \(t\geq1\), the logical bitstrings in the current global-best Ritz state were assigned to the nearest preceding reference by Manhattan distance, and their squared-coefficient-weighted mean within cluster \(\kappa\) defined \(\mathbf r_\kappa^{(t)}\).
Before each subsequent iteration, the cluster labels of samples in each fixed base pool were updated according to the same nearest-reference rule, while the pool itself remained unchanged.
For iteration-wise encoded expansion, this reassignment rule was instead applied to the original encoded samples before the recovery procedure was used to regenerate the logical base pool.

At iteration \(t\), mutant generation used the iteration-\((t-1)\) references.
For the encoded-expansion controls, the score for flipping position \(i\) was \(n_{\kappa,2i-1}^{(t-1)}\) when \(x_i=0\) and \(n_{\kappa,2i}^{(t-1)}\) when \(x_i=1\).
For unencoded expansion, the corresponding score was \(r_{\kappa,i}^{(t-1)}\) when \(x_i=0\) and \(1-r_{\kappa,i}^{(t-1)}\) when \(x_i=1\).
Whenever mutants were required, one fresh single-bit mutant and one fresh double-bit mutant were generated from each weighted base sample, and both inherited its weight.
The single-bit position was sampled in proportion to the flip scores, whereas the two distinct positions of the double-bit mutant were sampled sequentially without replacement using the same scores.

In the first iteration, no mutants were generated when the base states and \(\mathcal B_{\rm fix}\) were sufficient to fill \(D_{\rm tar}\).
When they were insufficient, every batch included all of them, and its remaining slots were sampled from a mixture of the single-mutant pool and the double-mutant pool, with the weights in each pool multiplied by \(1/2\).

At every later iteration, a fresh single-mutant pool and a fresh double-mutant pool were generated using the flip-score rules described above.
The candidate pool was formed from a mixture of the base pool, the single-mutant pool, and the double-mutant pool, with the weights in each pool multiplied by \(1/3\).
Carry-over states were not used as mutation sources, and mutants from earlier iterations were retained only if selected for carry-over.

The iterative uniform-random-recovery control and the three expansion controls used \(D_{\rm tar}\in\{0.5,1.0,1.5,2.0,2.5\}\times10^6\), five algorithmic seeds, \(N_B=10\), \(\epsilon_{\rm carry}=10^{-3}\), and \(\mathcal B_{\rm fix}=\{0^n,1^n\}\).
All four controls were run for exactly 40 iterations without early stopping.
Unless otherwise stated, all other algorithmic details were the same as in the main code-space recovery experiments.

\subsection{Software and computational environment}
Quantum-circuit construction, transpilation, and hardware execution on the IBM Heron r2 backend \texttt{ibm\_kingston} used Qiskit~2.2.3 and \texttt{qiskit-ibm-runtime}~0.44.0~\cite{qiskit}.
Measurement mitigation, BMM clustering, projected diagonalization, and DMRG calculations were performed using \texttt{mthree}~3.0.0~\cite{m3}, StepMix~3.0.0~\cite{stepmix}, PRIMME~3.2.3~\cite{primme}, and \texttt{quimb}~1.14.0~\cite{quimb}, respectively.
Major classical post-processing calculations were performed on an Amazon Web Services (AWS) EC2 c8i.48xlarge instance.
ChatGPT was used for language editing and coding assistance. 
All resulting text and code were reviewed and verified by the authors.

\vspace{0.3cm}

\backmatter

\bmhead{Data availability}
The processed encoded and unencoded hardware measurement datasets used in the analysis, together with the numerical results underlying the figures and tables in the main text and Supplementary Information, are available on Zenodo at \url{https://doi.org/10.5281/zenodo.22015275}.

\vspace{0.1cm}

\bmhead{Code availability}
The maintained source code for code-space recovery is available at \url{https://github.com/Byeongyong-Park/code-space-recovery-for-sqd}.
A versioned archival snapshot of the research code and production experiment notebooks supporting the results reported in this study, including the implementations and scientific settings used for the additional control experiments, is archived on Zenodo at \url{https://doi.org/10.5281/zenodo.22015275}.

\vspace{0.1cm}

\bmhead{Acknowledgements}
This work was supported by the National Research Foundation of Korea (NRF) grants funded by the Korean government (MSIT) (nos. RS-2025-00519428 and RS-2026-25612745) and by an Institute of Information \& Communications Technology Planning \& Evaluation (IITP) grant funded by the Korean government (MSIT) (no. RS-2025-25464788).
The authors also acknowledge the Urban Big Data and AI Institute of the University of Seoul (\url{http://ubai.uos.ac.kr}) for providing computing resources and Singularity Quantum Inc. for providing access to AWS computing resources.
D.A. also acknowledges the National Quantum Laboratory (QLab) at the University of Maryland and the Cleveland Clinic Quantum Innovation Catalyzer Program for providing supportive research environments.

\vspace{0.1cm}

\bmhead{Author contributions}
B.P. conceived the study, developed the methodology, performed the theoretical analysis, and designed the numerical experiments.
B.P. and S.K. implemented the software and conducted the numerical experiments.
B.P., S.K., D.A., and K.J. analyzed and interpreted the results. 
K.J. and D.A. acquired funding. 
K.J. supervised the project. 
B.P. wrote the initial manuscript draft, and all authors reviewed and edited the manuscript.


\bmhead{Competing interests}
D.A. is affiliated with Singularity Quantum Inc., which provided access to AWS cloud-computing resources used in this study. 
The other authors declare no competing interests.

\bibliography{sn-bibliography}

\end{document}